\documentclass[doublecol]{epl2}
\usepackage{graphicx}% Include figure files
\usepackage{dcolumn}% Align table columns on decimal point
\usepackage{bm}% bold math
\usepackage{bbm}

\title{Finite-Temperature Atomic Structure of $\bf{180^{\circ}}$ Ferroelectric Domain Walls in PbTiO$_3$}% Force line breaks with \\

\author{Arzhang Angoshtari\inst{1} \and Arash Yavari \inst{1}}

\institute{ \inst{1} School of Civil and Environmental Engineering,
  Georgia Institute of Technology, Atlanta, GA 30332.
}%


\date{\today}% It is always \today, today,
\abstract{In this letter we obtain the finite-temperature structure
of $180^{\circ}$ domain walls in PbTiO$_3$ using a quasi-harmonic
lattice dynamics approach. We obtain the temperature dependence of
the atomic structure of domain walls from $0 K$ up to room
temperature. We also show that both Pb-centered and Ti-centered
$180^{\circ}$ domain walls are thicker at room temperature; domain
wall thickness at $T=300 K$ is about three times larger than that of
$T=0 K$. Our calculations show that Ti-centered domain walls have a
lower free energy than Pb-centered domain walls and hence are more
likely to be seen at finite temperatures.}

\begin{document}

\maketitle

\section{Introduction}

Ferroelectric perovskites have been the focus of intense research in
recent years because of their potential applications in high strain
actuators, high density storage devices, etc.
\cite{BhattacharyaRavichandran2003}. It is known that macroscopic
properties of ferroelectrics strongly depend on domain walls, which
are extended two-dimensional defects. Any fundamental understanding
of ferroelectricity in perovskites requires a detailed understanding
of domain walls in the nanoscale (see \cite{DawberRabeScott2005} and
references therein). Theoretical studies of domain walls have
revealed many of their interesting features. From both theoretical
calculations and experimental works, it has been observed that the
thickness of domain walls can vary from thin walls, which consist of
only a few atomic spaces to thick walls, which are in the order of a
few micrometers. There have been studies using ab inito calculations
\cite{MeyerVanderbilt2001, Padilla1996, Poy1999} and anaharmonic
lattice statics calculations \cite{YaOrBh2006b} suggesting that
ferroelectric domain walls are atomically sharp. Hlinka and Marton
\cite{HlinkaMarton2006} analyzed $90^{\circ}$ domain walls in
BaTiO$_3$-like crystals in the framework of the phenomenological
Ginzburg-Landau-Devonshire (GLD) model and obtained domain wall
thickness of $3.6~nm$ at room temperature. Chrosch and Salje
\cite{ChroschSalje2006} measured the domain wall thickness in
LaAlO$_3$ in the temperature range $295-900~K$ by x-ray diffraction.
They observed that domain wall thickness increases from about
$20~{\AA}$ to $200~{\AA}$ and that the variation of domain wall
thickness with temperature is linear at low temperatures. Using
scanning probe microscopy, Iwata \textit{et al} \cite{Iwata2003}
found complex $180^{\circ}$ domain walls with thicknesses $1-2~\mu
m$ in PZN-$20\%$PT. Lehnen \textit{et al} \cite{Lehnen2000}
investigated $180^{\circ}$ domain walls in PbTiO$_{3}$ using
electrostatic force microscopy (EFM) and piezoelectric force
microscopy (PFM) and observed thick $180^{\circ}$
domain walls at room temperature with thickness of about $5~\mu m$. Shilo
\textit{et al} \cite{Shilo2004} studied the structure of
$90^{\circ}$ domain walls in PbTiO$_3$ by measuring the surface
profile close to emerging domain walls and then fitting it to the
soliton-type solution of GLD theory. Using this technique they
observed that the domain wall thickness is about $1.5~nm$ but with a
wide scatter. They suggested that the presence of point defects
within the domain wall is responsible for such variations. Lee \textit{et
al} \cite{Lee2005} provided a model to investigate the effect of
point defects on the domain wall thickness. See also
\cite{AngYavari2010} for a similar study.

Domain walls have been studied using different techniques in the
atomic scale at $T=0K$. However, one would be interested to know how
different the structure and thickness of a $180^{\circ}$ domain wall
at room temperature are compared to those at $T=0K$. In this letter,
we study the structure of Pb and Ti-centered $180^{\circ}$ domain
walls in PbTiO$_3$ as a function of temperature in some detail. We
first start with the static configuration of domain walls and
iteratively optimize the free energy for a small temperature, e.g.
$T_1=5K$. The optimized structure at $T_1$ will be the reference
configuration for a higher temperature $T_2=T_1+\Delta T$.
Continuing in this way we optimize the structure of the domain wall
up to $T=300K$. This temperature range is where quantum effects are
important and hence molecular dynamics simulations can not be used.
This is also the temperature range where quasi-harmonic
approximation is reasonable.

\begin{figure*}[]
\centerline{\hbox{\includegraphics[scale=0.9,angle=0]
{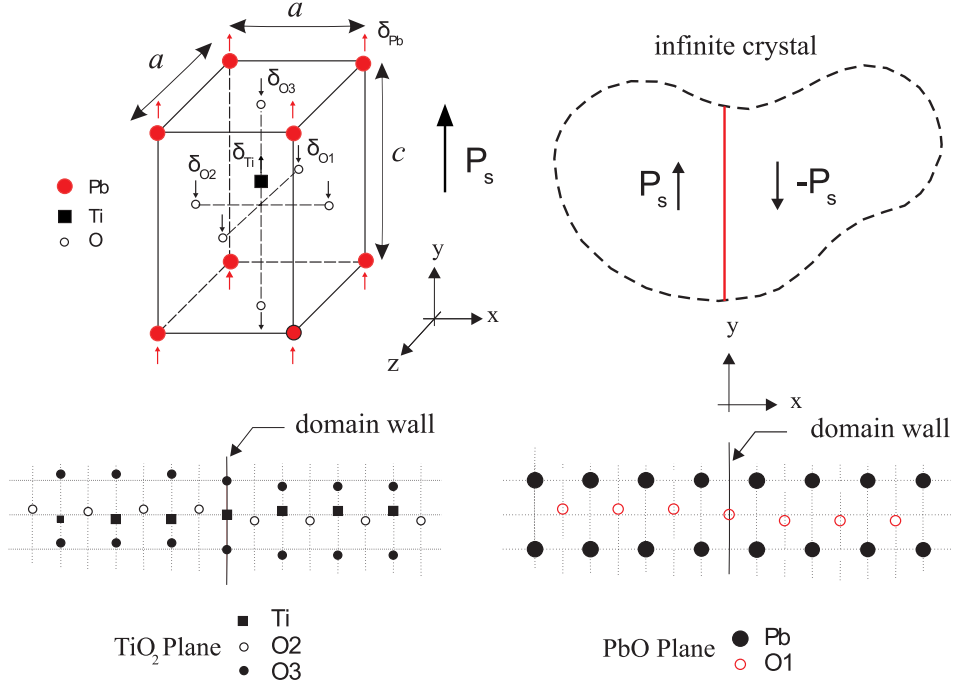}
                 }}
\caption{Reference configuration of a Ti-centered $180^{\circ}$
domain wall shown in both the TiO$_2$ and PbO planes. Note that
cores and shells on the domain wall have no relative shifts and
cores and shells on the left and right sides of the wall have
opposite relative shifts. \label{Reference}}
\end{figure*}

\begin{figure*}[t]
\centerline{\hbox{\includegraphics[scale=0.75,angle=0]{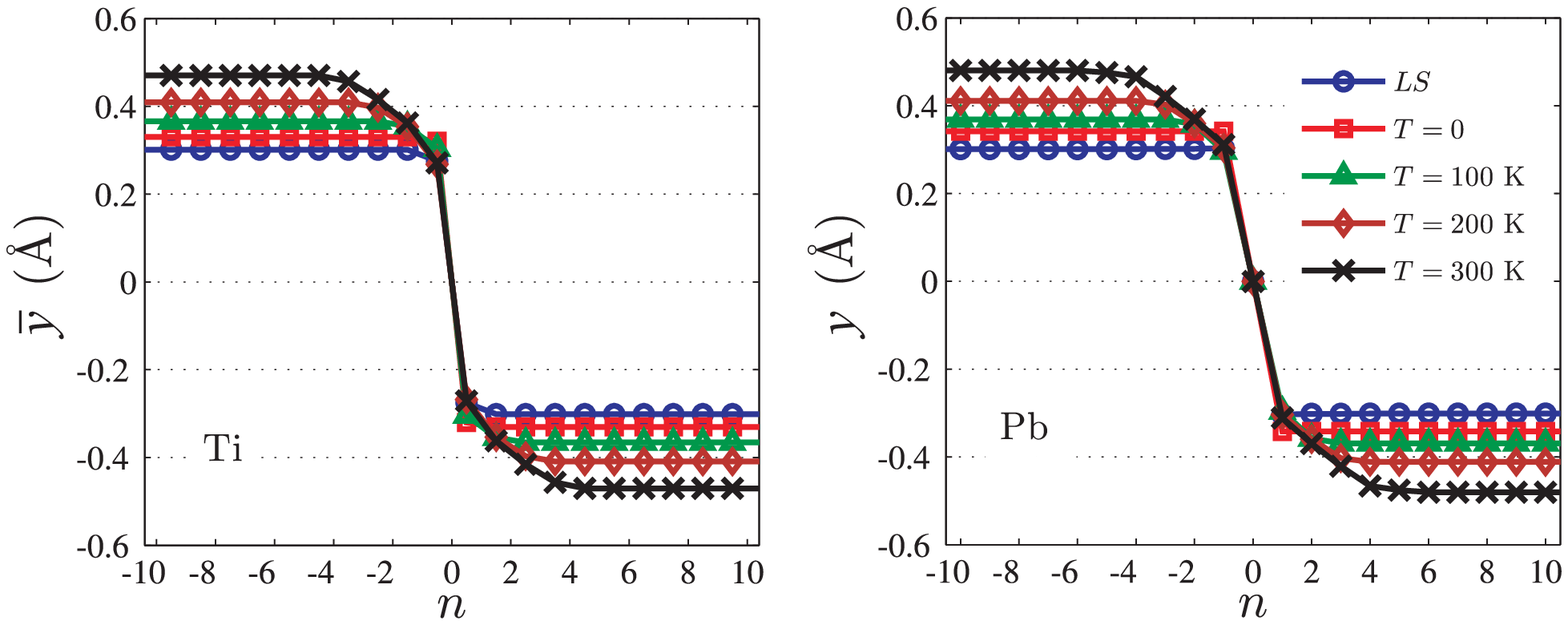}
\vspace*{-0.4in}
                 }}
\caption{The y-coordinates of Pb and Ti cores in the Pb-centered
$180^{\circ}$ domain wall as a function of temperature. $LS$ denotes
the lattice statics solution. \label{NP_PbTi}}
\end{figure*}

\begin{figure*}[t]
\centerline{\hbox{\includegraphics[scale=0.75,angle=0]{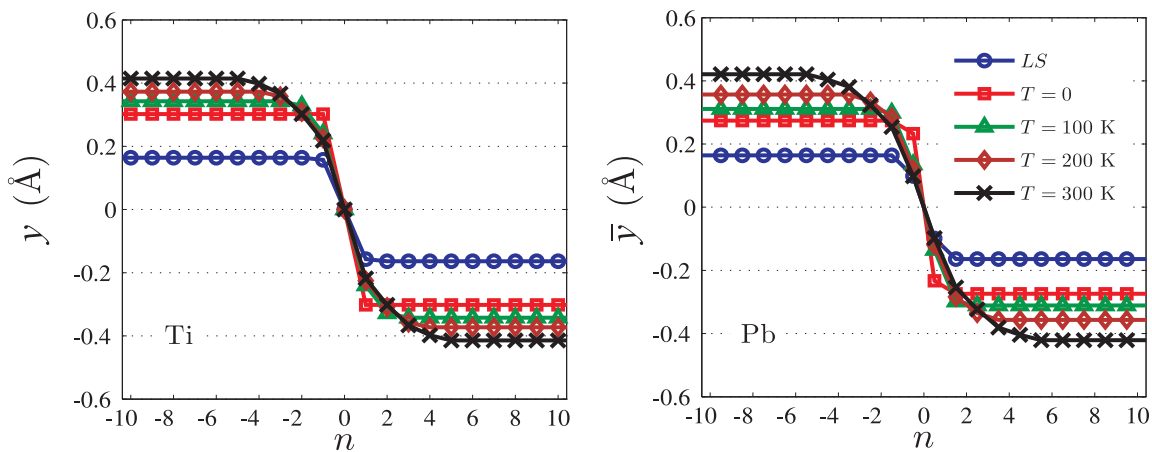}
\vspace*{-0.4in}
                 }}
\caption{The y-coordinates of Pb and Ti cores in the Ti-centered
$180^{\circ}$ domain wall as a function of temperature. $LS$ denotes
the lattice statics solution. \label{NT_PbTi}}
\end{figure*}

\section{Method of Calculation}The geometry of a Ti-centered $180^{\circ}$ domain wall is shown in
Fig. \ref{Reference}. The domain wall is in a $(100)$-plane and the
polarization vector is in the y-direction and changes by $180$
degrees across the domain wall. In this geometry, depending on which
cations are placed on the domain wall, two types of domain walls are
possible: Pb-centered and Ti-centered. In this work, we consider
both types and calculate their finite temperature structures. For
our calculations, we use the shell potential developed by Asthagiri
\textit{et al} \cite{Asthagiri2006} for PbTiO$_3$. In this
potential, each ion is described by a core and a massless shell. The
short range interactions between Pb-O, Ti-O and O-O shells are
described by the Rydberg potential $V_{sr}(r)=(A+Br)\exp(-r/C)$,
where A, B and C are potential parameters.  The cores and shells of
each ion have Coulombic interactions with cores and shells of all
the other ions. The core and shell of an atom interact by an
anharmonic spring of the form $V_{cs}(r)=(1/2)k_2 r^2 +(1/24)k_4
r^4$, where $k_2$ and $k_4$ are constants.

The structure of a domain wall is calculated by an analytical free
energy optimization method. This method was developed by
Kantorovich \cite{Kantorovich1995} and has been applied to various different
systems, e.g. in
\cite{Kantorovich1995II,Gale98,TaylorBarreraAllanBarron1997}. We
refer the reader to these references for the complete details of the
method. A domain wall is an example of a defective lattice with a
1-D symmetry reduction \cite{YaOrBh2006a,KavianpourYavari2009}. Here
we briefly explain the free energy optimization method exploiting
symmetry reduction. For more details see
\cite{YavariAngoshtari2009}.

The multi-lattice of PbTiO$_3$ can be partitioned into the union of
some 2-D equivalence classes that are parallel to the domain wall.
Therefore, we reduce the dimension of the problem from three to one,
i.e., we can write $\mathcal{L}=\bigsqcup_{I}^{}\bigsqcup_{\alpha\in
\mathbbm{Z}}\mathcal{L}_{I\alpha}$, where $\mathcal{L}$,
$\mathcal{L}_{I\alpha}$, and $\mathbbm{Z}$ are the lattice, 2-D equivalence classes,
and the set of integers, respectively. Note that $j=J\beta$ means
that the atom $j$ is in the $\beta$\emph{th} equivalence class of
the $J$\emph{th} sublattice \cite{YaOrBh2006a}. As an approximation
similar to that of \cite{LesarNajafabadiSrolovitz1989}, we assume
only a finite number of neighboring equivalence classes, $N$, on
each side of the domain wall  and assume the temperature-dependent
bulk configuration outside this region. Therefore, the size of the
effective dynamical matrix would be $15\times 2N$. To obtain the
lattice structure around the domain wall at a finite temperature, we
minimize the Helmholtz free energy $\mathcal{F}$, calculated based
on the quasi-harmonic approximation with respect to the
configuration $\left\{\mathbf{X}^j\right\}_{j\in\mathcal{L}}$ at a
finite temperature $T$.  Note that away from the domain wall the two
half lattices approach their temperature-dependent configurations.
We can write the free energy of the defective lattice
$\mathcal{F}\equiv\mathcal{F}\left(\{\mathbf{X}^j\}_{j\in\mathcal{L}},T\right)$,
as

%-----------------------------
\begin{eqnarray}
    \mathcal{F} \!&=&\!
    \mathcal{E}\left(\{\mathbf{X}^j\}_{j\in\mathcal{L}}\right)
    \nonumber \\
         &+&\!\! \sum_{\mathbf{k}}\sum_{i}^{}\!\left\{\frac{1}{2}\hbar
    \omega_i(\mathbf{k})
    \!+\! k_BT\ln \! \left[\! 1-\exp\left(\!-\frac{\hbar
    \omega_i(\mathbf{k})}{k_BT}\right)\right]\right\}, \nonumber \\
\end{eqnarray}
%-----------------------------
where $\mathcal{E}$ is the total static energy of the lattice and
$\omega(\mathbf{k})$ is the frequency at wave number
$\mathbf{k}\in\textsf{B}$ with $\textsf{B}$ the first Brillouin
zone of the sublattices.  Note that for calculating the derivatives
of the frequencies, we exploit the analytical method of
Kantorovich \cite{Kantorovich1995}.

For optimization of the free energy, we use the quasi-Newton method
with the Broyden-Fletcher-Goldfurb-Shanno update for calculating the
approximate inverse Hessian in each step \cite{PressTVF1989}. We
should mention that to converge to the optimized configuration, one
should select an initial configuration close to the solution. Thus,
for calculating a finite temperature configuration, we start with a
nominal configuration (see Fig. \ref{Reference}), which is obtained
by relaxing the bulk at temperature $T=0K$, and then using this bulk
configuration with opposite directions of core-shell shifts on the
two sides of the $180^{\circ}$ domain wall. Then, we relax the
nominal configuration and obtain the lattice statics solution. Next,
using the lattice statics solution, we obtain the lattice
configuration at zero temperature and then we use temperature steps
of $\Delta T = 25K$, and obtain the optimized configuration at a given
finite temperature. This way we observe that the quasi-Newton method
converges relatively fast. Assuming force tolerance of $0.05~
\textrm{eV/{\AA}}$, our solutions converged after about $20$ to $40$
iterations depending on the temperature. Note that we only assume
periodicity of unit cells in the y- and z-directions. In the
x-direction we assume $N$ unit cells in each side of the domain wall
and use the temperature-dependent bulk configurations as the far
field boundary conditions. Our numerical experiments show that $N=12$
would be enough to capture the atomic structure near the domain walls
up to $T=300K$ as we do not see changes in the structure by choosing
larger $N$.  In our calculations, we used a $1 \times 3\times 3$
\textit{k}-point Monkhorst-Pack mesh \cite{MonkhorstPack1976}. Also
for calculating the classical Coulombic potential and force, we used
the damped Wolf method \cite{Wolf99}.

\section{Numerical Results for $\mathbf{180^{\circ}}$ Domain Walls}

Because displacements of core and shell of the same atom are close,
we only report the core displacements. Figure \ref{NP_PbTi} shows
the y-coordinates (tetragonal coordinates) of the Pb and Ti cores
relative to a Pb core on the Pb-centered domain wall. Figure
\ref{NT_PbTi} shows the y-coordinates of the Pb and Ti cores
relative to a Ti core on the Ti-centered domain wall.  In these
figures $\bar{y}=y-(c/2)$, where $c$ is the temperature-dependent
lattice parameter in the tetragonal direction. The lattice
parameters change with temperature such that by increasing
temperature, unit cells transform from tetragonal to cubic
\cite{BeharaHinojosa2008}. Here to calculate lattice parameters at
a finite temperature, we separately optimized the bulk lattice at that temperature. Note that far from the domain wall, each half lattice
approaches its corresponding temperature-dependent bulk
configuration.

In these figures, $LS$ denotes the lattice statics solution (static
energy minimization). We observe that the lattice statics solution
and the configuration obtained by the free energy minimization at
zero temperature although predicting nearly the same domain wall
thicknesses, are different due to the zero-point motions; the
lattice statics method ignores the quantum effects. It is known that
zero-point motions can have significant effects in some systems
\cite{KohanoffAndreoniParrinello1992}. Here we observe that
zero-point motions affect the Ti-centered domain wall more than the
Pb-centered domain wall; zero-point motions change the lattice
statics solutions by about $15\%$ in the Pb-centered domain wall and
by about $50\%$ in the Ti-centered domain wall. Since the atomic
displacements normal to the domain wall are small (they are of order
$10^{-2}$ {\AA}) we do not report them here. However, we will comment on them in the sequel.

Increasing the temperature from $0$ to $300K$, we observe that the
domain wall thickness increases from $1 nm$ to about $3 nm$. This
qualitatively agrees with experimental observations for $90^{\circ}$
domain walls in PbTiO$_3$ by Foeth \textit{et al}
\cite{FoethStadelmannRobert2007} and domain walls in LaAlO$_3$ by
Chrosch and Salje \cite{ChroschSalje2006} who observed that domain
wall thickness increases with temperature. They measured an average
domain wall thickness from room temperature up to the Curie
temperature. It is worth mentioning that from
Ginzburg-Landau-Devonshire theory, domain wall thickness is
proportional to $|T-T_c|^{-1}$ \cite{ChroschSalje2006}, where $T_c$
is the Curie temperature. This means that for low temperatures
domain wall thickness is linear in temperature. We observe this
linear behavior in our numerical simulations. We should also mention
that a similar trend was observed in a lattice of dipoles
\cite{YavariAngoshtari2009}.

Note also that domain wall
thickness cannot be defined uniquely very much like boundary layer
thickness in fluid mechanics. Here, domain wall thickness is by
definition the region that is affected by the domain wall, i.e.
those layers of atoms that are distorted. One can use definitions
like the $99\%$-thickness in fluid mechanics and define the domain
wall thickness as the length of the region that has $99\%$ of the
far field rigid translation displacement. What is important is that
no matter what definition is chosen, domain wall ``thickness"
increases by increasing temperature.

Recently, it has been observed that there may be local normal and
transverse polarizations near domain walls. For example,
Goncalves-Ferreira \textit{et al} \cite{Goncalves2008} observed
local polarizations near domain wall of CaTiO$_3$ (nonpolar
material) parallel and perpendicular to the wall plane. In our
simulations, for both Pb and Ti-centered domain walls we observe
that polarization in c-direction switches within a few lattice
spacings in the vicinity of the domain wall and the polarization
normal to the domain wall is about $2\%$ of the polarization in the
c-direction. In particular, we see that normal displacements are
$5.0\%$ and $8.0\%$ of their corresponding c-displacements in the Pb
and Ti-centered domain walls, respectively. This agrees with the
results of Lee, et al. \cite{Lee2009} who observed non-zero
displacements normal to the domain wall. In their calculations,
polarization normal to the domain wall in the Pb and Ti-centered
$180^{\circ}$ domain walls are $2.5\%$ and $1.75\%$, respectively,
of the bulk polarization.

%-----------------------------
\begin{figure}[t]
\begin{center}
\includegraphics[scale=0.6,angle=0]{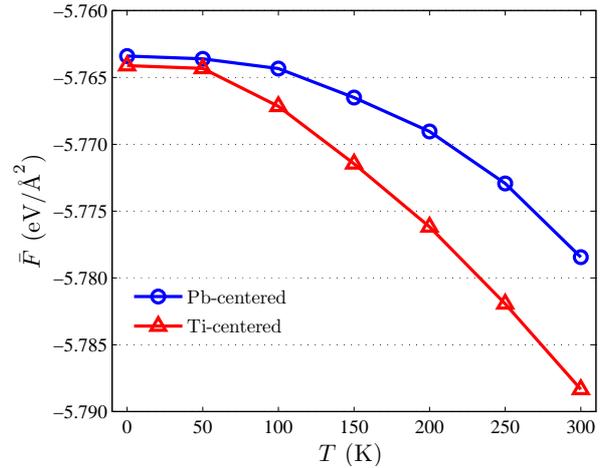}
\end{center}\vspace*{-0.2in}
\caption{ Free energy of the domain walls as a function of
temperature.} \label{free}
\end{figure}
%-----------------------------
%-----------------------------

In Fig. \ref{free}, we have plotted the free energy per unit cell, $\bar{F}=F/(N\times
a\times c)$, where $N$ is the number of relaxed unit cells and $a$
and $c$ are temperature-dependent lattice parameters, for the two
types of domain walls. In agreement with Meyer and Vanderbilt \cite
{MeyerVanderbilt2001}, we observe that Ti-centered domain walls have
a higher static energy, however we see that they have a lower free
energy and hence are the preferred domain wall configuration at
finite temperatures.

\section{Concluding Remarks}

In this work we obtained the finite-temperature structure of Pb and
Ti-centered $180^{\circ}$ domain walls in PbTiO$_3$ using a
quasi-harmonic lattice dynamics method. Our numerical results are in
good agreement with experimental measurements. We observed a strong
dependence of structure on temperature. In particular, $180^{\circ}$
domain walls at $T=300K$ are three times thicker than those at
$T=0K$. We also observed that free energy is a decreasing function
of temperature and free energy of a Ti-centered domain wall is
always lower than that of a Pb-centered domain wall and hence
Ti-centered domain walls are more likely to be seen at finite
temperatures.

\acknowledgments We thank an anonymous referee for useful comments.

\end{document}